\documentclass[]{spie}  %>>> use for US letter paper
\usepackage{amsmath,amsfonts,amssymb}
\usepackage{graphicx}
\usepackage[colorlinks=true, allcolors=blue]{hyperref}

\usepackage{epstopdf}

\title{Crowdsourcing for Identification of Polyp-Free \\ Segments in Virtual Colonoscopy Videos}

\author[a]{Ji Hwan Park}
\author[a]{Seyedkoosha Mirhosseini}
\author[a]{Saad Nadeem}
\author[a]{Joseph Marino}
\author[a]{\\Arie Kaufman}
\author[b]{Kevin Baker}
\author[b]{Matthew Barish}
\affil[a]{Department of Computer Science, Stony Brook University, Stony Brook, NY, USA}
\affil[b]{School of Medicine, Stony Brook University, Stony Brook, NY, USA}

\authorinfo{Further author information:\\
E-mail J.H.P., K.M., S.N., J.M., A.K.: \{ jihwpark $|$ smirhosseini $|$ sanadeem $|$ jmarino $|$ ari \}@cs.stonybrook.edu\\
E-mail K.B., M.B.: \{ Kevin.Baker $|$ Matthew.Barish \}@stonybrookmedicine.edu}

\begin{document} 
\maketitle

\begin{abstract}
Virtual colonoscopy (VC) allows a physician to virtually navigate within a reconstructed 3D colon model searching for colorectal polyps. Though VC is widely recognized as a highly sensitive and specific test for identifying polyps, one limitation is the reading time, which can take over 30 minutes per patient. Large amounts of the colon are often devoid of polyps, and a way of identifying these polyp-free segments could be of valuable use in reducing the required reading time for the interrogating radiologist. To this end, we have tested the ability of the collective crowd intelligence of non-expert workers to identify polyp candidates and polyp-free regions. We presented twenty short videos flying through a segment of a virtual colon to each worker, and the crowd was asked to determine whether or not a possible polyp was observed within that video segment. We evaluated our framework on Amazon Mechanical Turk and found that the crowd was able to achieve a sensitivity of 80.0\% and specificity of 86.5\% in identifying video segments which contained a clinically proven polyp. Since each polyp appeared in multiple consecutive segments, all polyps were in fact identified. Using the crowd results as a first pass, 80\% of the video segments could in theory be skipped by the radiologist, equating to a significant time savings and enabling more VC examinations to be performed.
\end{abstract}


\section{INTRODUCTION}
\label{sec:intro}  % \label{} allows reference to this section

Virtual colonoscopy (VC) employs CT scanning of a patient's abdomen and advanced visualization techniques to virtually navigate within a reconstructed 3D colon model searching for colorectal polyps, the precursor of cancer.  VC is widely recognized as a highly sensitive and specific test for identifying polyps in the colon~\cite{Johnson:2008}, and it is a comfortable, inexpensive, very low risk, and fast procedure.  However, several impediments still remain for widespread adoption of VC as a colorectal screening tool.  One of the major issues with VC is the large amount of interpretation time required by the radiologists.  In order to ensure full coverage of the colon surface, a typical protocol is to acquire two scans of the patient in different positions (e.g., supine and prone) and to then virtually traverse each scan in both antegrade and retrograde directions.  These multiple fly-throughs are used to ensure sufficient coverage of the colon surface, but are tedious and time consuming, as most regions of the colon are free of polyps.  This required reading time has reduced the capacity of a radiologist to perform a large number of VC readings.

Computer-aided detection (CAD) algorithms have been developed to identify polyps based on features such as shape, curvature, and wall thickness.  Current validated CAD algorithms have been shown to have false negative findings, missing lesions with adherent contrast medium, flat adenomas, and adenomas located on or adjacent to normal colonic folds~\cite{Summers:2008}.  In order to increase sensitivity, specificity is usually sacrificed, thus increasing the number of false positive polyp candidates that must be rejected.  This requires significant human intervention and time even though the majority of false positive candidates can be rejected even by individuals with minimal training. 

Crowdsourcing seeks to engage the general masses in innovative ways and solicit their inputs in solving diverse problems, and previous studies have shown that most workers are forthright in their intentions~\cite{Suri:2011}.  There have been a few attempts to combine CAD with crowdsourcing to reduce false positives~\cite{McKenna:2012,Wang:2011}.   These initial attempts have shown that the crowd can be used to reject false positive polyp candidates, however little work has been done to determine if the crowd can be used for primary polyp detection or as a means of excluding the presence of a polyp within a segment of colon.  Other work has shown that the performance of the crowd was roughly the same as that of a CAD system, but that workers who participated in two studies improved significantly in performance between the first and second study~\cite{nguyen:2012}.  Crowdsourcing has also been used to generate reference correspondences in endoscopic images~\cite{Maier-Hein2014a} and for performing instrument segmentation in laparoscopic images~\cite{Maier-Hein2014b}.

The ability to accurately determine polyp-free segments is what has the greatest chance to reduce reading time.  Since humans are extremely adept at recognizing shapes, we seek to use crowdsourcing to engage the general masses in identifying whether video segments from a VC fly-through contain a polyp or are polyp-free.  The purpose of this study is to determine whether participants can identify, with reasonable accuracy, whether or not a fly-through video of a segment of the human colon contains a possible polyp.  From this, we seek to observe if segments can be accurately identified as lacking possible polyps.  Based on the results we gather, we hope to further refine our system such that a first pass screening by the crowd will enable certain portions of the colon to be skipped or reviewed more quickly by the radiologist, enabling a faster read time for each VC case.

\section{METHODS}

\subsection{Endoluminal Videos}

In clinical VC systems, the standard view is a virtual fly-through of a volume rendered colon model which mimics the general appearance of a traditional optical colonoscopy.  This imagery presents a detailed, accurate, and understandable reconstruction of a view inside the colon.  As a first pass in screening, the radiologist will typically fly through the colon using this endoluminal view, examining the reconstructed mucosa for any polyps, which appear as protrusions.  Since this endoluminal view is the standard view in VC and the view most easily understood for laypeople, we render such views as the type of imagery to present to the crowd workers.  In order to maintain simplicity for the non-expert workers, other views that are available in the clinical VC system, such as CT slices and an overview mesh model, were not presented to the crowd.  Since we are interested in reducing the time required for the first pass screening, these additional views are unnecessary for the task at hand.

To create this endoluminal view, a virtual camera is placed along a path, and raycasting from the camera position through the CT volume results in the final generated imagery. 
Typically, a centerline is calculated through the colon lumen and is used as the flight path, yielding good general coverage.  However, due to the haustral folds and bends of the colon, some portions will be missed~\cite{hong:2007:spie}, necessitating both antegrade and retrograde fly-throughs of each colon model.

For this research, we have used the commercial FDA approved Viatronix V3D\textsuperscript{\textregistered}-Colon virtual colonoscopy system~\cite{viatronix} to generate the endoluminal videos, which are captured at 15 frames per second (fps) with a resolution of $256 \times 256$ pixels.  After the fly-throughs of the entire colon models were established and videos captured, the journey through the colon was divided into approximate equi-time segments of twelve seconds in length.  These segments have slight overlaps of two seconds to ensure that a polyp which might be located on a video boundary is not missed.  To ensure coverage of the majority of the colon, the fly-throughs for each dataset were performed in both antegrade and retrograde directions for both the supine and prone scans for each patient.

When generating the fly-through videos, it is possible to adjust some parameters that are used for the rendering.  One parameter is the flight speed, allowing users to go slower or faster based on their experience.  While a faster speed will require less reading time, it can also lead to areas being overlooked as they appear only briefly.  A second parameter is the field of view; a wide angle \emph{fisheye} view with a 120 degree viewing angle can be enabled to allow for greater coverage of the colon wall.  While more of the wall becomes visible in a single fly-through, this option introduces significant distortion, which can cause confusion for the viewer.  Prior to this work, we performed a preliminary study using videos with the fisheye view on (for increased coverage) and at a moderate speed.  After receiving user feedback on those videos, we discovered that the distortion from the fisheye view, especially around bends, was confusing to the users.  A number of users also complained that the speed was too fast.  Incorporating this feedback, our present work makes use of videos rendered with a normal 90 degree viewing angle and at a slower fly-through rate of about half the speed of the previous videos.

\begin{figure}[t]
\centering
\includegraphics[width=13cm]{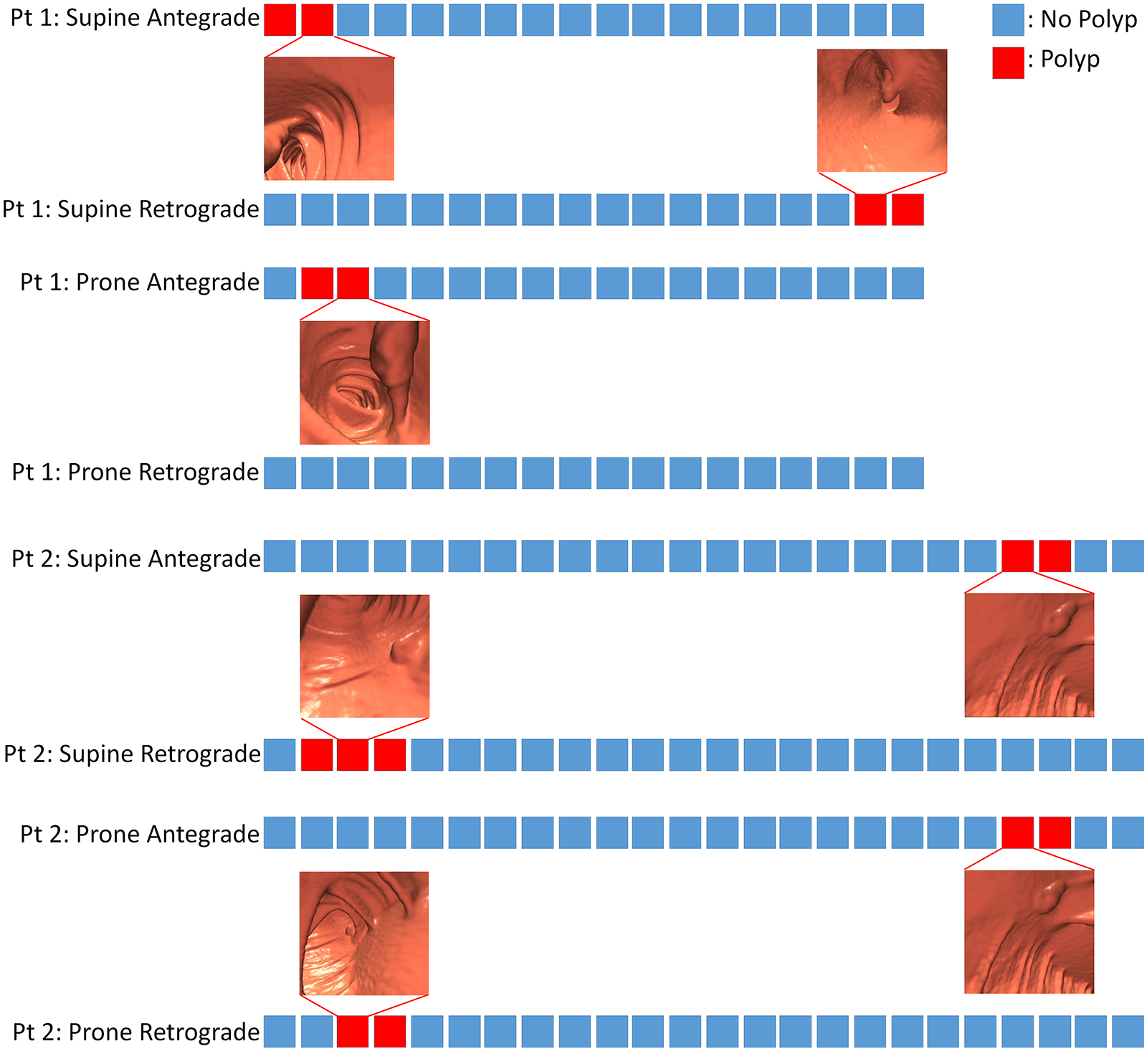}
\caption{Timeline view of the videos, with each box representing a video segment from the entire video.  The blue boxes represent segments which contain no polyps, while the red boxes represent segments that contain a clinically proven polyp.  Note that the polyps always appear in at least two consecutive segments.  Sample frames from the videos are also shown illustrating the polyps as they appear in the videos.}
\label{fig:timeline}
\end{figure}

\subsection{Data and Ground Truth}

The data used was captured from two patients, with a total of eight complete fly-through videos being recorded.  These include antegrade and retrograde navigation in both the supine and prone datasets for each patient.  These eight full length videos were divided into a total of 163 video segments, yielding an average of approximately 20 video segments per complete fly-through video.

The colon CT data which was used for this study had previously undergone VC examination, with each patient found to have one polyp which was clinically confirmed by OC.  We use these clinically proven polyp locations as the ground trurth, and note at what time during the full video fly-throughs each polyp would come into view and then exit the view.  These notations were used when splitting the videos to maintain the ground truth of which video segments contained a polyp.

The timelines of the eight videos and their corresponding segments are shown in Figure~\ref{fig:timeline}.  It can be noted that for each polyp's appearance in a full-length video, it would  appear in at least two consecutive video segments due to the slow navigation speed.  Therefore, it is imperative that the crowd identify the polyp in at least one video segment in which it is present, but not necessarily in all video segments in which it appears.  Since each polyp appears in at least two consecutive videos, identification of at least one video segment by the crowd as containing a polyp candidate is sufficient to ensure that the polyp would be identified during further interrogation by a radiologist.  The Patient 1 retrograde videos also illustrate the importance of making use of both antegrade and retrograde fly-throughs, as the polyp which was visible in the other three fly-throughs was not visible in the retrograde fly-through of the prone dataset.

\subsection{User Interface}

To present our study to the non-expert crowd workers, we made use of the Amazon Mechanical Turk (MTurk) platform, which has become popular recently as a way of obtaining reliable crowd workers at modest cost while allowing us to reach a large and diverse population of users~\cite{paolacci:2010}.  Our study was identified on the MTurk platform as an academic survey which would require the labeling of videos and was limited only to users in the United States.

When users first select this task, they are provided with brief directions about the purpose of the system as well as two short tutorial videos.  These tutorial videos each contain a polyp, which is annotated in order to illustrate to the workers what they should be looking for.  Although our interest lies in identifying polyp-free segments, it is more natural to ask workers to look for an object, rather than for the absence of the object.  To this end, the users are requested to identify if each video segment contains a polyp candidate.  Since they are not presented with the more advanced tools required to make a final determination, the term polyp candidate is used instead of polyp.  In relation to our desired results, this also indicates that we expect a not insignificant number of false positive results for areas which cannot be further determined from the fly-through video alone.

After viewing the brief tutorial, the user is provided sequentially with twenty video segments randomly selected from the pool of 163 video segments.  For each video, there is a selection for \emph{Yes} (a polyp candidate is present in the current video) or \emph{No} (there is no polyp candidate present in the current video).  After selecting one of the options, the user submits the result and continues directly to the next video.  The videos start playing automatically when the page loads, and the user can pause the video, replay the video, and drag the time slider as desired to view specific frames of interest.  Once the user has clicked \emph{Submit}, the user cannot go back to a previous video.  After the user has inspected all twenty videos, comments solicited, which we will use to further refine our system in the future.

\begin{figure} [t]
\footnotesize
\begin{center}
\setlength{\tabcolsep}{0pt}
\begin{tabular}{ccc} %% tabular useful for creating an array of images 
\includegraphics[width=.49\columnwidth]{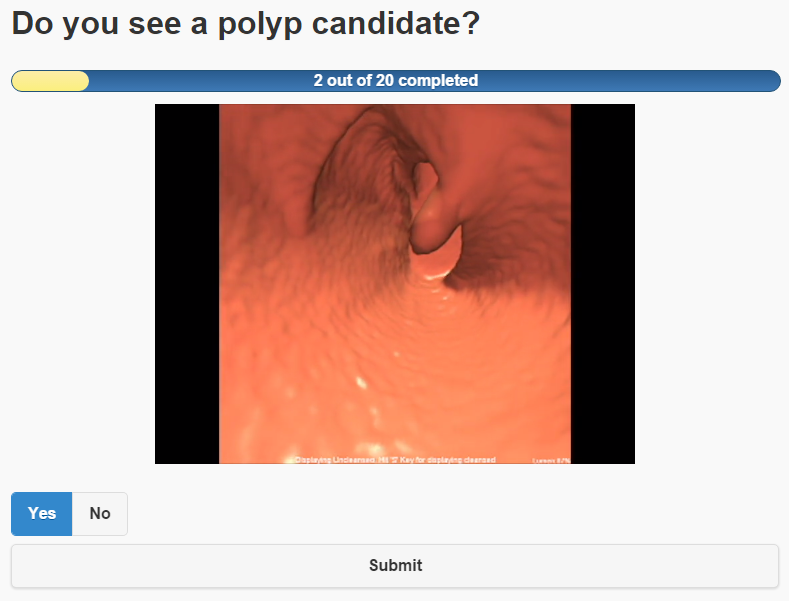}& \hspace{2.5mm} &\includegraphics[width=.49\columnwidth]{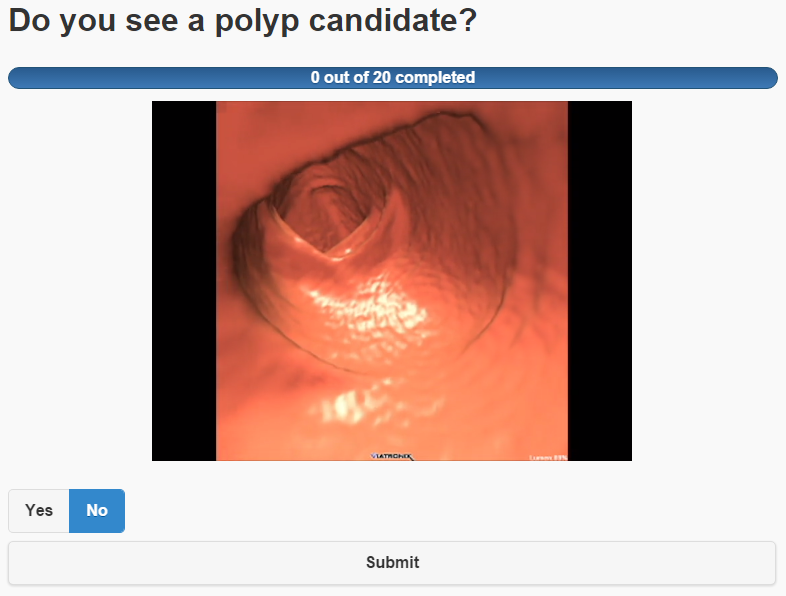}\\
(a)& &(b)
\vspace{-2mm}
\end{tabular}
\end{center}
\caption{\label{fig:ui} Our interface showing a VC frame (a) with a polyp candidate in the view and (b) with no polyp in the view.}
\end{figure}

The presentation of the videos was randomized between participants, and each video was approximately twelve seconds in length.  Each user was allowed to perform the task only once.  Examples of our interface are shown in Figure~\ref{fig:ui}, with one view displaying a polyp candidate and one with no polyp candidate.  The bar at the top of the screen allows the users to keep track of their progress through the twenty videos.

\section{RESULTS}

A total of 144 participants each viewed a random selection of 20 video segments from the pool of 163 total video segments, yielding a total of 2,880 determinations.  Of these determinations, 436 were true positives, 551 were false positives, 1,684 were true negatives, and 209 were false negatives.  The selection of the individual video segments for presentation to the crowd was randomized, and each video segment was viewed, on average, by approximately 18 workers.  Out of the total pool of 163 video segments, there were 15 video segments which contained a clinically proven polyp and 148 video segments which did not.  To obtain a crowd consensus for each video segment, a segment was denoted as containing a possible polyp if at least half of the respondents for that video indicated \emph{Yes} (i.e., the segment does contain a polyp).  Using this metric to analyze the results for each video segment yielded 12 true positives, 20 false positives, 128 true negatives, and 3 false negatives, giving a sensitivity of 80.0\% and a specificity of 86.5\%.  These results are summarized in Table~\ref{tab:results}.

Although the per-video segment results are of interest, of greater interest are the per-polyp results for each full-length video.  Some polyps only appear very briefly in the first or last few frames of a video segment, and are thus difficult to identify in the confines of that segment.  However, due to the fly-through speed, each polyp appears in at least two consecutive video segments.  Since polyps will occur in multiple video segments, we determine that identification by the crowd of a polyp in at least one of the video segments  in which it is contained is sufficient for the polyp as a whole to be identified.  Due to this, while the sensitivity was only 80.0\% based on per-video results, each ground truth polyp in each full-length fly-through was in fact identified.  The detailed results for the 15 videos containing polyps are shown in Table~\ref{tab:results-polyps}.

We also evaluated the results of a VC-trained radiologist who examined all 163 videos, the determinations of which yielded 13 true positives, 19 false positives, 129 true negatives, and 2 false negatives, giving a sensitivity of 86.7\% and a specificity of 87.2\%, results which were only slightly better than the crowd.  These results are summarized in Table~\ref{tab:results}.  As with the crowd results, although the radiologist had two false negatives, all actual polyps were identified in at least one video segment.

\begin{table}[t]
\center
\caption{Results of the crowd and a trained radiologist in determining which video segments contained a polyp or did not contain a polyp.  Summarized are the number of true positives (TP), number of false positives (FP), number of true negatives (TN), number of false negatives (FN), the sensitivity (Sens.), and the specificity (Spec.).}
\vspace{1mm}
\label{tab:results}
\begin{tabular}{| l | c | c | c | c | c | c |}
  \hline
  & TP & FP & TN & FN & Sens. & Spec. \\
  \hline
  Crowd Workers & 12 & 20 & 128 & 3 & 80.0\% & 86.5\% \\
  \hline
  Radiologist & 13 & 19 & 129 & 2 & 86.7\% & 87.2\% \\
  \hline
\end{tabular}
\end{table}

\begin{table}[t]
\center
\caption{The video segments containing a ground truth polyp and the crowd results.  For each segment's row the following information is given: the number of crowd users that responded \emph{Yes}; the number that responded \emph{No}; the overall crowd determination for that video segment, being a true positive (TP) or false negative (FN); and whether or not that polyp was identified in either that video segment or a neighboring segment is given.}
\vspace{1mm}
\label{tab:results-polyps}
\begin{tabular}{| l | r | r | c | l |}
  \hline
  Video Segment & \emph{Yes} & \emph{No} & Result & Polyp Identified? \\
  \hline
  Patient 1, Supine, Antegrade, Segment 1 & 27 & 15 & TP & Yes, in this segment\\
  Patient 1, Supine, Antegrade, Segment 2 & 9 & 32 & FN & Yes, in segment 1\\
  \hline
  Patient 1, Supine, Retrograde, Segment 17 & 37 & 8 & TP & Yes, in this segment\\
  Patient 1, Supine, Retrograde, Segment 18 & 49 & 9 & TP & Yes, in this segment\\
  \hline
  Patient 1, Prone, Antegrade, Segment 2 & 0 & 15 & FN & Yes, in segment 3\\
  Patient 1, Prone, Antegrade, Segment 3 & 10 & 6 & TP & Yes, in this segment\\
  \hline
  Patient 2, Supine, Antegrade, Segment 21 & 45 & 5 & TP & Yes, in this segment\\
  Patient 2, Supine, Antegrade, Segment 22 & 41 & 8 & TP & Yes, in this segment\\
  \hline
  Patient 2, Supine, Retrograde, Segment 2 & 39 & 12 & TP & Yes, in this segment\\
  Patient 2, Supine, Retrograde, Segment 3 & 33 & 10 & TP & Yes, in this segment\\
  Patient 2, Supine, Retrograde, Segment 4 & 15 & 31 & FN & Yes, in segments 2 and 3\\
  \hline
  Patient 2, Prone, Antegrade, Segment 21 & 42 & 8 & TP & Yes, in this segment\\
  Patient 2, Prone, Antegrade, Segment 22 & 27 & 11 & TP & Yes, in this segment\\
  \hline
  Patient 2, Prone, Retrograde, Segment 3 & 24 & 24 & TP & Yes, in this segment\\
  Parient 2, Prone, Retrograde, Segment 4 & 38 & 15 & TP & Yes, in this segment\\
  \hline
\end{tabular}
\end{table}

\section{DISCUSSION}

To our knowledge, this is the first study to evaluate the ability of non-expert workers to identify whether or not a polyp candidate was present in a VC video, and thus identify polyp-free segments.  This study gives confidence that the crowd, as a whole, can achieve detection rates near that of an individual radiologist.  While the sensitivity on a per-video basis was 80.0\%, all polyps were identified in at least one video segment since each polyp appeared in at least two consecutive segments.  In comparison, a VC-trained radiologist achieved results on a per-segment basis with only one fewer false negative.

With 131 segments identified by the crowd as being devoid of polyp candidates, this represents 80\% of the video segments we presented.  Since every polyp from each full fly-through was identified by the crowd in at least one video segment, these 131 segments could be skipped, or reviewed more quickly, by the radiologist in a system which incorporated the crowd results.  This represents a significant time savings, and would allow an individual radiologist to review a greater number of cases.

Our results also indicate that this type of task is well suited for crowdsourcing.  From the informal comments gathered after the task, a number of users commented that they found the task to be interesting.  A few noted difficulties, and we believe that more training for the user could help to overcome this and lead to even better results.  A significant improvement might also be possible by allowing the same user to participate multiple times (with different videos), allowing the user to learn more with each task.  We are also considering the possibility of seeding the videos with some \emph{a priori} data for which we already know the results.  Using this data, we could then remove low performing workers from the consensus results, improving the overall results.  In the future, we would also like to compare the crowd as a whole against a cohort of VC-trained radiologists.

We also plan to examine the possibilities of combining crowdsourcing with other technologies for further improvement.  For example, the crowd results might be useful for reducing false positives from CAD, or as a method of augmenting a CAD system~\cite{hong:2006:tvcg}.  Furthermore, registering the supine and prone colons~\cite{nadeem:2017:tvcg} would allow for results from all four fly-throughs (antegrade and retrograde in both supine and prone) to be combined for greater accuracy.

\section{CONCLUSIONS}

We have presented a brief crowdsourcing study in which workers were tasked with identifying whether or not a short VC video segment contains a polyp.  We found that the crowd achieved a sensitivity of 80.0\% and specificity of 86.5\% per video segment, compared to a sensitivity of 86.7\% and specificity of 87.2\% by a trained radiologist.  However, since each polyp would appear in at least two consecutive video segments taken from the entire VC fly-through, each polyp was in fact identified by the crowd.  Future work will investigate how to best incorporate the crowd results into a VC system to allow the physician to perform the examination more quickly with the same, or possibly greater, accuracy.

\acknowledgments % equivalent to \section*{ACKNOWLEDGMENTS}       
 
This work has been partially supported by the National Science Foundation grants IIS0916235, IIP1069147, and CNS0959979.

% References
\bibliography{spie17_crowdsourcing_vc} % bibliography data in report.bib
\bibliographystyle{spiebib} % makes bibtex use spiebib.bst

\end{document}